\documentclass[aps,prl]{revtex4}

\begin{document}

\begin{flushright}
{\it DOE/ER/40762-269\\
UM PP 03-023}\\
\end{flushright}

\title{\large \bf  Electromagnetic Properties of
the $\Delta$  in the Large $N_c$ and Chiral Limits}

\vspace{15mm}

\author{\bf Thomas D. Cohen}

\affiliation{ Department of Physics,
University of Maryland,\\
College Park, MD 20742-4111,USA \vspace{.15in} }

\begin{abstract}

 Relations between nucleon,
$\Delta$ and nucleon-$\Delta$ electromagnetic properties
associated with chiral physics which have been derived in a pure
large-$N_c$ limit are greatly modified in a combined large $N_c$
and chiral limits; the large $N_c$ relations are valid for
$(M_\Delta-M_N)/m_\pi <<1$.   The E2/M1 ratio for the
nucleon-$\Delta$ transition is shown to scale as $\epsilon^{-1}$
(with $\epsilon\sim 1/N_c$ , $\epsilon \sim m_\pi/\Lambda $ with
$\Lambda$ a typical hadronic scale) in the combined large $N_c$
and chiral expansions as opposed to $N_c^{-2}$ in the pure large
$N_c$ expansion.

\end{abstract}


\maketitle

It has been known for some time that contracted $\rm{SU}(2N_f)$
spin-flavor symmetry of baryons
\cite{GerSak83,GerSak84,DasMan93a,DasMan93b,DasJenMan93,DasJenMan94,Jen93}
which follows from large $N_c$ QCD\cite{Hoo74,Wit79} relates
various $\Delta$ and $\Delta$-nucleon transition matrix elements
to nucleon properties.  These relations are model independent in
the large $N_c$ limit; they are fixed entirely from the
symmetry.  Subleading effects in $1/N_c$ are not generally fixed.
However for certain properties the leading $1/N_c$ correction
vanishes.  For example, the nucleon-$\Delta$ transition magnetic
moment \cite{DasJenMan94} is  fixed relative to the nucleon
isovector moment up to relative order $1/N_c^2$.

Recently the contracted ${\rm SU}(2N_f)$ symmetry was used in an
elegant way to relate other electromagnetic properties of the
$\Delta$ and the N-$\Delta$
transition\cite{BucLeb00,BucHesLeb02,BucLeb02}, including the
electric charge radius of the $\Delta$\cite{BucLeb00,BucLeb02},
the quadrupole moment of the $\Delta$\cite{BucHesLeb02} and the
nucleon-$\Delta$ quadrupole transition
moment\cite{BucHesLeb02,BucLeb02}.  In a related work, the small
value in E2/M1 electromagnetic decays of the $\Delta$ quadrupole
transition moment to the magnetic transition moment) is explained
qualitatively due to large $N_c$ scaling behavior (which is shown
to go as $1/N_c^2$) \cite{JenJiMan02}. The scaling rules in
ref.~\cite{JenJiMan02} and the relations in
refs.~\cite{BucLeb00,BucHesLeb02,BucLeb02} are all model
independent in the large $N_c$ limit of QCD.
 To the extent that the physical world is
close to the large $N_c$ limit, one might expect the relations in
refs.~\cite{BucLeb00,BucHesLeb02,BucLeb02} to hold to reasonable
approximation in the real world and the qualitative explanation
in ref.~\cite{JenJiMan02} to be valid.

However,  while the $1/N_c$ expansion may work well for generic
nucleon or $\Delta$ properties, it is certainly possible that for
certain observables, the model-dependent correction terms may be
so large  as to render the $1/N_c$ expansion  invalid.
 The present work argues on the basis of chiral considerations
that one expects the electromagnetic quantities discussed in
refs.~\cite{BucLeb00,BucHesLeb02,BucLeb02,JenJiMan02} to have
large corrections (excepting those relations which depend solely
on isospin).   These corrections are large enough in the real
world to overwhelm the large $N_c$ suppression factors for the
Delta charge radius and quadrupole moment. The issue of the
quadrupole transition moment is more subtle: the ratio E2/M1 is
of order $1/N_c^2$; however, the coefficient multiplying the
$1/N_c^2$  is chirally enhanced and, hence, is parametrically
large. Thus, the empirical smallness of this ratio depends in
large part on fortuitous numerical factors.

One expects  correction terms to the leading large $N_c$
relations to be anomalously large if the quantity in question is
dominated by large distances. The longest distance contributions
to baryon structure comes from virtual pions which are
anomalously light  because the pions are pseudo-Goldstone bosons.
Chiral perturbation theory \cite{BerKiaMei95} explains why such
quantities are singular in the chiral limit.  The singular
behavior occurs due to infrared enhancements in pion loops which
dominate the infrared because the pion is the only small
excitation scale in the problem. As noted in
refs.~\cite{CohBro92,Coh95a,Coh95b}, however, this no longer true
in the large $N_c$ limit since the nucleon-$\Delta$ mass
splitting scales as $1/N_c$ \cite{DasMan93a,DasJenMan93}.

The results of refs.~\cite{CohBro92,Coh95a,Coh95b} may be
summarized as follows: i) Static nucleon properties which are
singular in the chiral limit do not have a uniform combined
chiral and large $N_c$ limit, {\it i.e.} the chiral limit and
large $N_c$ limits do not commute.  ii)  There exists a
well-defined limiting procedure for these quantities in which
$m_\pi \rightarrow 0 \; N_c \rightarrow \infty$ with $N_c m_\pi$
fixed. iii) The leading singular behavior for these properties in
this combined limit can be calculated from a one pion loop
calculation in a generalization of chiral perturbation theory in
which the $\Delta$ is kept as an explicit light degree of freedom
and in which the pion-baryon coupling respect the contracted SU(4)
symmetry of refs.
\cite{GerSak83,GerSak84,DasMan93a,DasMan93b,DasJenMan93,DasJenMan94,Jen93}.
iv)  The Skyrme model\cite{BroZah86} and other hedgehog soliton
models \cite{Bir90} when treated using semiclassical quantization
reproduce all of the results for these chirally singular
quantities provided that these quantities are computed with the
large $N_c$ limit taken prior to the chiral limit.

These results imply that a typical static nucleon quantity in this
category (which will be denoted here as $\Theta$) can be written
as
\begin{equation}
\Theta = \frac{a  \, g_A^2}{f_\pi^2 m_\pi} \, f_\Theta \left(
\Delta/m_\pi \right ) + {\rm subleading \, \,
terms}\label{typical}
\end{equation}
where $a$ is a numerical constant and $\Delta \equiv M_\Delta -
M_N$; the functional form of $f_\Theta$ depends on the quantum
numbers of $\Theta$ and the number of pion propagators which
contribute \cite{explain}; the subleading terms are less singular
than the first term in the limit $m_\pi \rightarrow 0 \; N_c
\rightarrow \infty$ with $N_c m_\pi$ fixed. The function
$f_\Theta$ has the property,
\begin{equation}
f_\Theta (0) =1 \; \; \; {\rm while} \; \; \;  f_\Theta (x) < 1 \;
\; {\rm for} \; \; x>0 \label{f}  \; .
\end{equation}
The various parameters scale with $N_c$ as $ {g_A^2}/{f_\pi^2}
\sim N_c$  and $ \Delta \sim 1/N_c$. The large $N_c$ limit taken
prior to the chiral limit yields $f_\Theta(0)=1$ while the
opposite ordering of limits give $f_\Theta(\infty ) <1$, clearly
demonstrating the noncommutation of the limits.  The coefficient
$a$ can be found directly from the Skyrme model  by matching the
Skyrme model prediction for $\Theta$ with eq.~(\ref{typical})
while setting $f_\theta$ to unity.

It is useful to establish a power counting scheme around the
combined limit.  This can be done in a manner analogous to that
 in the ``small scale'' expansion scheme which generalizes
chiral perturbation theory to include explicit $\Delta$ degrees of
freedom \cite{HemHolKam98,JenMan91}.  In the present context if
one denotes the small counting parameter as $\epsilon$ then
\begin{equation}
1/N_c \sim \epsilon \; \; {\rm and} \; \; m_\pi / \Lambda \sim
\epsilon \label{count}
\end{equation}
where $\Lambda$ is a typical hadronic scale.  This power counting was
 constructed so that the
leading power of a divergence in $\epsilon^{-1}$ does not depend
on how one approaches the combined large $N_c$ and chiral
limits.  For example $\Theta$ in eq.~(\ref{typical}) will scale
as $\epsilon^{-2}$ regardless of the value of the ratio of
$\Delta$ to $m_\pi$.  This property allows one to extract the
leading divergence with $\epsilon$ directly from the Skyrme model
(which gives the correct behavior in the pure large $N_c$ limit).

The analysis of refs.~\cite{CohBro92,Coh95a,Coh95b} can be
extended to the study of $\Delta$ properties and properties of the
nucleon-$\Delta$ transition in a rather straightforward manner as
the arguments apply to all baryons in the $I=J$ tower. The
function $f_\Theta$ which controls the interpolation between the
two limits now can depend on the quantum numbers of the baryons
(i.e. nucleon vs. $\Delta$) and is not restricted to be less than
unity. Again, the scaling behavior of an observable in the
combined limit can be extracted from the Skyrme model.  Here we
foucus on the electric charge radii of the nucleon and $\Delta$,
the nucleon-$\Delta$ electric quadrupole transition moment and
the electric quadrupole moment of the $\Delta$. These are easily
computed in any Skyrme type model with semiclassical quantization
at leading order in the $1/N_c$ expansion \cite{CohBro92}:
\begin{eqnarray}
\langle r^2 \rangle_{\Delta; m_I} \, & = & \, \frac{1}{2} \langle
r^2 \rangle_{I=0} \, + \,  m_I \langle r^2 \rangle_{I=1} \label{rsqD}\\
\langle r^2 \rangle_{N; m_I} \, & = & \, \frac{1}{2}\langle r^2
\rangle_{I=0} \, + \,  m_I \langle r^2 \rangle_{I=1}  \label{rsqn}\\
\langle J m m_I | Q_\mu | J' m' {m'}_I \rangle \, & = & \, \langle
J|| Q || J' \rangle \, m_I \delta_{m, {m'}_I} \frac{\langle J m 2
\mu |
J' m' \rangle}{\sqrt{2 J'+1}} \label{WE} \\
\langle 3/2 || Q || 1/2 \rangle \, & = &\,- \frac{\sqrt{50}}{8} \,
\langle 3/2 || Q || 3/2 \rangle \,= \, - \sqrt{125} \, Q \label{qme} \\
{\rm with} \; \; Q & \equiv & \frac{\sqrt{8}}{5} \, \langle
\Delta^{+} \, m=1/2| Q_0 | {\rm p } \, m=1/2 \rangle \label{Qdef}
\end{eqnarray}
where the results hold to leading order in the $1/N_c$
expansion.  The values of $Q$, $ \langle r^2 \rangle_{I=1}$ and
$\langle r^2 \rangle_{I=0}$ all depend on the details of the
model but all scale as $N_c^0$.  Equation (\ref{WE}) simply
reflects the structure of the Wigner-Eckert theorem along with
the large $N_c$ result that the quadrupole operator at leading
order is purely isovector.  Note that eq.~(\ref{qme}) relates the
nucleon-$\Delta$ electric quadrupole transition to the static
quadrupole moment of the $\Delta$.

The relations in eqs.~(\ref{qme}) and (\ref{Qdef}) are equivalent
to the leading order model independent results of
refs.~\cite{BucHesLeb02} while  eqs.~(\ref{rsqD}) and(\ref{rsqn})
encode the leading order relations of ref. \cite{BucLeb00}.  Note
also, that the conclusion of ref.~\cite{JenJiMan02} that the
E2/M1 ratio scales as $N_c^{-2}$ depends on the fact that the the
nucleon-$\Delta$ electric quadruole matrix element scales as
$N_c^0$; a result seen in Skyrme type models as seen above.

As $m_\pi \rightarrow 0$, the classical pion tail in the Skyrme
profile becomes increasingly long-ranged and dominates the
contributions to both the isovector charge radius and the
electric quadrupole properties.  As has long been
known\cite{CohBro86}, the leading chiral behavior for both sets
of quantities in these models can be directly related to the
strength of the pion tail and thus to $g_A$ leading to additional
relations:
\begin{equation}
 \langle r^2 \rangle_{I=1} \, = \, \sqrt{500 \pi} \,  Q \, = \,
 \frac{5 \, g_A^2 \Delta}{16 \pi f_\pi^2 \,  m_\pi} \; \label{prop}.
 \end{equation}
Note that both quantities diverge as $1/m_\pi$ in the  chiral
limit and the arguments of refs.~\cite{CohBro92,Coh95a,Coh95b}
apply.  In terms of the power counting of  eq.~(\ref{count}) they
scale as,
\begin{equation}
Q \sim \epsilon^{-1} \; \;  \; \; \; \langle r^2 \rangle_{I=1}
\sim \epsilon^{-1} \; . \label{epsscal}\end{equation} Since the
scaling with $\epsilon$ does not depend on the ordering of limits,
the scaling in eq.~(\ref{epsscal}) holds even away from the pure
large $N_c$ limit.

Equation (\ref{epsscal}) is significant in the context of E2/M1
in $\Delta$ decays.  This ratio was shown to go as $N_c^{-2}$ in
ref.~\cite{JenJiMan02} with one factor of $1/N_c$ coming from
kinematics while the other came from the fact that $Q \sim N_c^0$
and the magnetic transition moment, $\mu$, scaled as $N_c^{-1}$.
However, in the combined chiral and large $N_c$ counting scheme,
$Q \sim \epsilon^{-1}$ and $\mu \sim \epsilon^{-1}$.  Thus, in
the combined counting E2/M1 $\sim \epsilon$.  There is one less
power of suppression in the combined counting than in the pure
$1/N_c$ counting due to a chiral enhancement.  Thus it is hard to
argue that one understands the very small empirical E2/M1 ratio
purely from scaling arguments.  It is plausible that smallness is
associated with the large numerical coefficients in the
proportionality  constant relating the electric quadrupole
transition to the isovector charge radius in eq.~(\ref{prop}).
These coefficients come from the detailed structure of the group
theory and not from scaling behavior with $m_\pi$ or $N_c$.

The quantities of interest diverge in the chiral limit and so the
relations in eqs.~(\ref{rsqD}),(\ref{rsqn}) and (\ref{qme}). are
modified as one approaches the combined large $N_c$ and chiral
limits along the lines noted in
refs.~\cite{CohBro92,Coh95a,Coh95b}.  They can be expressed as:
\begin{eqnarray}
\langle r^2 \rangle_{\Delta; m_I} & = &
 m_I \, \langle r^2 \rangle_{I=1}^{\rm LN } \, \,
 f_{\Delta,\Delta, r^2}(\Delta/m_\pi) + {\cal O}(\epsilon^0)
 \label{rsqDm}\\
\langle  r^2 \rangle_{N; m_I} & = & \,  m_I \, \langle r^2
\rangle_{I=1}^{\rm LN} \, \,
f_{N, N, r^2}(\Delta/m_\pi) + {\cal O}(\epsilon^0) \label{rsqnm}\\
\langle 3/2 || Q || 1/2 \rangle & = & - \sqrt{125} \, \,  Q^{\rm
LN} \, \,
 f_{\Delta,N,Q}(\Delta/m_\pi) + {\cal O}(\epsilon^0) \label{qmem} \\
\langle 3/2 || Q || 3/2 \rangle & = & \sqrt{20} \, \,  Q^{\rm LN}
\, \,  f_{\Delta,\Delta,Q}(\Delta/m_\pi) + {\cal O}(\epsilon^0)
\label{q2mem}
\end{eqnarray}
$\langle r^2 \rangle_{I=1}^{\rm LN }$ and $Q^{\rm LN}$ are the
expressions for  $\langle r^2 \rangle_{I=1}$ and $Q$ when the
large $N_c$ limit is taken prior to the chiral limit. As such
they are calculable in the Skyrme model and are given in
eq.~(\ref{prop}).  The functions are
$f_{B,B',\Theta}(\Delta/m_\pi)$ where $B$ and $B'$ are the
initial and final baryon states and $\Theta$ denotes the quantity
in question. These functions are calculable from one loop pion
graphs.  The detailed form of these will be given in a subsequent
publication.

Equations (\ref{rsqDm}), (\ref{rsqnm}), (\ref{qmem}) and
(\ref{q2mem}) imply large corrections to the model-independent
large $N_c$ predictions of refs.~\cite{BucLeb00,BucHesLeb02} due
to chiral enhancements.  In the large $N_c$ limit, the $f$
functions are unity while for generic values of $\Delta/m_\pi$
they differ from one another by amounts of order unity .  This
spoils the large $N_c$ proportionality rules of
refs.~\cite{BucLeb00,BucHesLeb02}  relating $\Delta$ charge radii
to nucleon charge radii and the $\Delta$ quadrupole moments to
the nucleon-$\Delta$ transition moment.  The relations depending
solely on isospin are, of course, still valid.

In fact, the situation may be worse than this for the real world
value of $\Delta/m_\pi \sim 2$. The analytic structure of the
functions $f_{B,B',\Theta}(\Delta/m_\pi)(\Delta/m_\pi)$ is
nontrivial. When $B$ and $B'$ are nucleons, the functions are
analytic for all values $\Delta/m_\pi$ as can been from their
explicit forms as given in refs.~\cite{CohBro92,Coh95a,Coh95b}.
This reflects the lack of open channels. Similarly, when  $B$ and
$B'$ are either both $\Delta$ states or one nucleon and one
$\Delta$, they will be analytic for values of $\Delta/m_\pi <1 $
as there are no open channels.   However, for $\Delta/m_\pi \le
1$ the pion-nucleon channel is open.  This raises a conceptual
problem in the analysis since the analysis in
refs.~\cite{CohBro92,Coh95a,Coh95b} is strictly valid for only
static quantities and it is unclear how to define static
quantities of the $\Delta$ when it is coupled to an open channel.
Mathematically, this will be reflected by nonanalytic behavior in
$f_{B,B',\Theta}(\Delta/m_\pi)(\Delta/m_\pi)$.  If one continues
the forms of the $f_{B,B',\Theta}$ to the regime where
$\Delta/m_\pi < 1$, one will  encounter imaginary parts.  These
represent the ambiguity in defining the values of the quantity
for the $\Delta$ resonance given the need to separate the
resonance contribution  from a continuum background. Such a
separation is always model dependent. Provided the imaginary part
is small, however, this ambiguity will be small and meaningful
extraction of $\Delta$ properties are possible. For the case of
the electric quadrupole transition from nucleon to $\Delta$ there
will be an imaginary part, but there is no reason to expect it to
dominate.  Calculations of E2/M1 for the $\Delta$ nucleon
transition have significan imaginary parts but they do not
dominate \cite{ButSavSpring93}.

The case of the $\Delta$ charge radius and $\Delta$ quadrupole
moment are different; one expects very large imaginary parts in
the open-channel regime and thus it may be difficult to relate
these quantities to experimental observables. To see this,
consider the regime very close to threshold, {\it i.e.} where $
|1 - \Delta^2/m_\pi^2| << 1$. One can develop an effective theory
valid for this regime. The central physical idea is that below
threshold in this regime, the pion is effectively nonrelativistic
at long distances.  Thus, the long distance part of the $\Delta$
can be visualized as a nonrelativistic p-wave between a  pion and
a nucleon and as such it is described by a wave function which
goes as $\sim e^{- m_\pi \sqrt{1 - \Delta^2/m_\pi^2} \, r}/r^2$
at asymptotic distances. From this it is easy to deduce that in
the vicinity of $\Delta/m_\pi = 1$ the functions scale as
\begin{equation}
f_{\Delta, N, r^2} \,   =  \, \frac{a}{\sqrt{1 -
\frac{\Delta^2}{m_\pi^2}}} \, + \, {\rm non singular \, \, terms}
\; \; \; \; \; \; \; \;
 f_{\Delta,\Delta, Q} \,  =  \, \frac{b}{\sqrt{1 -
\frac{\Delta^2}{m_\pi^2}}} \, + \, {\rm non singular \, \,
terms}\label{nonan}
\end{equation}
where $a$ and $b$ are constants.

As  one approaches the threshold from the bound region, the
charge radius and the quadrupole moment both diverge.  As one
moves into the unbound region the functions are totally dominated
by their imaginary parts and as such the quantity becomes
ambiguous. One way to understand this ambiguity is in terms of
contributions from the background. The background is a p-wave
nucleon-pion continuum state which extends out to spatial
infinity. Thus, changes in the model-dependent separation between
resonance and background can move arbitrarily large contributions
to these quantites from the continuum  into the resonance.  In
this circumstance it is not clear how to extract the $\Delta$
resonance contribution for these quantities. As one continues
well above threshold and outside the region of this effective
theory it is remains unclear exactly how one should interpret the
$\Delta$ charge radius and quadrupole moment.   Reference
\cite{BucHesLeb02} notes the practical difficulties in extracting
the static quadrupole moment of the $\Delta$ from experiment.  It
remains an open question as to whether there exists continuations
of the expressions for these quantities which is both valid above
threshold and is meaningful in the sense that they can be related
to experimentally measurable quantities. Thus the interplay
between the chiral and large $N_c$ limits not only alters the
relations of refs.~\cite{BucLeb00,BucHesLeb02}, it also makes the
interpretation of the charge radius and quadrupole moment of the
$\Delta$ problematic in the real world where $\Delta/m_\pi>1$.

In summary, the interplay between the chiral and large $N_c$
limits implies that taking a combined limit greatly alters
relations between chirally sensitive nucleon, $\Delta$ and
nucleon-$\Delta$ electromagnetic properties such as charge radii
and quadrupole matrix elements from those which have been derived
previously in a pure large-$N_c$ limit. The  pure large $N_c$
relations become accurate when $\Delta/m_\pi <<1$ while in the
real world $\Delta/m_\pi \sim 2$. Moreover, the $\Delta$ charge
radius and electric quadrupole moment become ambiguous for
$\Delta/m_\pi >1$.   The E2/M1 ratio for the nucleon-$\Delta$
transition is shown to scale as $\epsilon$ in the combined large
$N_c$ and chiral expansions as opposed to $N_c^{-2}$ in the pure
large $N_c$ expansion.  It is less suppressed in the combined
limit making it more difficult to understand why the ratio is so
small.

This work is supported by the U.S.~Department of Energy grant
DE-FG02-93ER-40762. The author gratefully acknowledges useful
discussions with X. Ji, A Manohar, R. Lebed, and A. Buchmann.

\end{document}